\documentclass{article}

\PassOptionsToPackage{numbers, sort&compress}{natbib}

\usepackage[preprint]{neurips_2025}

\usepackage[utf8]{inputenc}
\usepackage[T1]{fontenc}
\usepackage{hyperref}
\hypersetup{colorlinks=true, linkcolor=red!60!black, citecolor=blue!60!black, urlcolor=magenta!60!black}
\usepackage{url}
\usepackage{amsmath}
\usepackage{amsfonts}
\usepackage{amssymb}
\usepackage{nicefrac}
\usepackage[linesnumbered,ruled]{algorithm2e}
\usepackage{graphicx}
\usepackage{subcaption}
\usepackage{multirow}
\usepackage{pifont}
\usepackage{booktabs}
\usepackage[table]{xcolor}
\usepackage{array}
\usepackage{makecell}
\usepackage{microtype}
\usepackage{placeins}


\title{MultiCard: \textbf{Multi}ple Attribute \textbf{Card}inality Estimation via Deep Learning}

\author{%
  Mobina Koochaknia \qquad Maryam Ramezani \qquad Mohammad Hasan Shafaat \\
  Sharif University of Technology, Tehran, Iran \\
  \texttt{\{mobina.kochaknia01, maryam.ramezani, mhmd.shafaat84\}@sharif.edu}
}

\begin{document}

\maketitle

\begin{abstract}
Cardinality estimation is essential for query optimization, but most existing learned estimators focus on a single predicate type, such as string LIKE clauses or numerical range filters. Despite this, analytical workloads in practice often merge both, but the accuracy of string estimators diminishes with longer patterns. We introduce MultiCard, a comprehensive, learned cardinality estimator. It unifies the modeling of LIKE predicates and numerical filters into a single end-to-end architecture. By combining a Transformer (for long-range, sparse patterns) and an LSTM (for short, frequent patterns), the proposed method achieves robust pattern modeling. Simultaneously, a numerical encoder combines catalog histograms and structural predicate features. This allows the model to dynamically assess the accuracy of histogram estimations across various operators and data skew conditions. A final Transformer layer captures interactions across heterogeneous predicates before an MLP predicts query cardinalities. We also introduce UPCI, a workload generator that produces length-balanced LIKE patterns, multi-column numerical predicates, and exact cardinality labels for training and evaluation. Experiments on five real-world datasets show that, on string-only workloads, the proposed method achieves the best or near-best Q-error across all datasets compared with state-of-the-art methods. On mixed multi-attribute queries, MultiCard reduces PostgreSQL’s mean Q-error by up to 16× while maintaining an inference latency below 5 ms.
\end{abstract}

\section{Introduction}

Cardinality estimation is a fundamental task in database management systems (DBMS). It involves predicting the number of rows a query will return prior to execution, allowing the query optimizer to choose an efficient plan without incurring actual execution costs. However, this estimation process is not uniform; different query types present distinct operational challenges and application requirements.

For instance, string search queries are heavily utilized in domains such as information retrieval, social media text filtering, DNA sequence alignment, and  data cleaning. The primary challenge in estimating these queries is the extreme variance in string lengths. Existing models often suffer from degraded accuracy when handling longer strings. The length of string predicates plays a significant role in cardinality estimation accuracy. Short patterns often produce high-cardinality results due to frequent occurrences, whereas longer substring predicates are typically more selective and sparse. As string length increases, the underlying distribution becomes more complex, making traditional histogram-based estimators less effective for accurately modeling textual correlations and substring frequencies. This is a critical bottleneck for realistic workloads, as text columns in real-world datasets frequently contain long strings with skewed, non-uniform distributions. In recent attempts to model the dependencies between words in a string, LSTMs have been widely employed, such as in the studies by CLIQUE \cite{kwon2025cardinality} and DREAM \cite{kwon2022cardinality}. Nevertheless, the efficiency of these methods degrades in the context of longer strings.

Queries containing numerical predicates are equally ubiquitous, appearing in statistical dashboards, time-series databases, IoT analytics, and financial systems. Because numerical range filters represent one of the most common database operations, classical estimators based on histograms and independence assumptions—were primarily designed around them. However, these assumptions break down when multiple numerical filters are combined, since column correlations are ignored, leading to compounding estimation errors as the number of predicates grows.

The core issue, however, is that real-world workloads rarely consist of isolated query types. Practical queries demand processing complex combinations of these conditions, encompassing both numerical predicates and string operations with highly variable lengths. \textcolor{black}{We refer to such queries as multi-attribute queries, as they combine predicates over multiple attributes, including both numerical and string attributes. Recent efforts have attempted to address combined predicates. For example, \cite{zeng2024price} deployed a model integrating join and numerical predicates. However, handling the full spectrum of mixed, real-world constraints remains an open problem.}
To address the complexities of real-world workloads, we propose a novel end-to-end architecture capable of simultaneously processing \texttt{LIKE} string predicates and numerical conditions on a single table. We overcome the existing challenges using MultiCard through following contributions:
\begin{itemize}
    \item \textbf{Unified multi-attribute estimation:} We propose MultiCard,  a unified learned estimator designed for muli-attribute queries involving both LIKE string predicates and numerical filters. Instead of estimating predicates individually and aggregating the results, MultiCard captures predicate interactions jointly through an output Transformer.

    \item \textbf{Length-aware string encoding:} We demonstrate that a single LSTM-based encoder’s effectiveness decreases as the length of LIKE patterns grows. Accordingly, we propose SPE, a hybrid string predicate encoder that integrates LSTM and Transformer representations via a length-aware mechanism to improve handling of both short frequent patterns and long sparse strings.

    \item \textbf{Histogram-based numerical encoding:} We introduce NPE, a numerical predicate encoder that combines catalog histogram data with predicate features. The model can refine the optimizer’s estimates through predicate operators and data distributions.

    \item \textbf{Workload generator:} We introduce UPCI, a workload generator designed to create length-balanced query workloads featuring LIKE predicates and multi-column numerical filters with precise cardinality labels. The generated workloads are publicly available to support future research.

    \item \textbf{Comprehensive assessment:} MultiCard is tested on five real-world datasets, measuring both accuracy and execution time. The results indicate that MultiCard greatly enhances cardinality estimation for multi-attribute queries, and the proposed SPE encoder delivers performance comparable to current string predicate estimators.
\end{itemize}
The source code, data, and other artifacts are available at \url{https://github.com/ramezani-lab/MultiCard}.



\section{Related Work}

Historically, traditional cardinality estimation relied heavily on statistical algorithms and data profiling techniques, such as histograms, random sampling, and sketches. Recently, some studies have revisited these foundations; for example, \cite{salihouglu2026cardinality} focuses on structural formulas and database statistics, providing a comprehensive mathematical and graphical framework for integrating and analyzing traditional data summarization methods. In contrast, modern learning-based cardinality estimation models are generally classified into three distinct approaches: data-driven, query-driven, and hybrid. A summary of these factors along with a feature-by-feature comparison of each approach can be found in Table \ref{comparefeature}.

\begin{figure}[htbp]
  \centering
  \includegraphics[width=\textwidth]{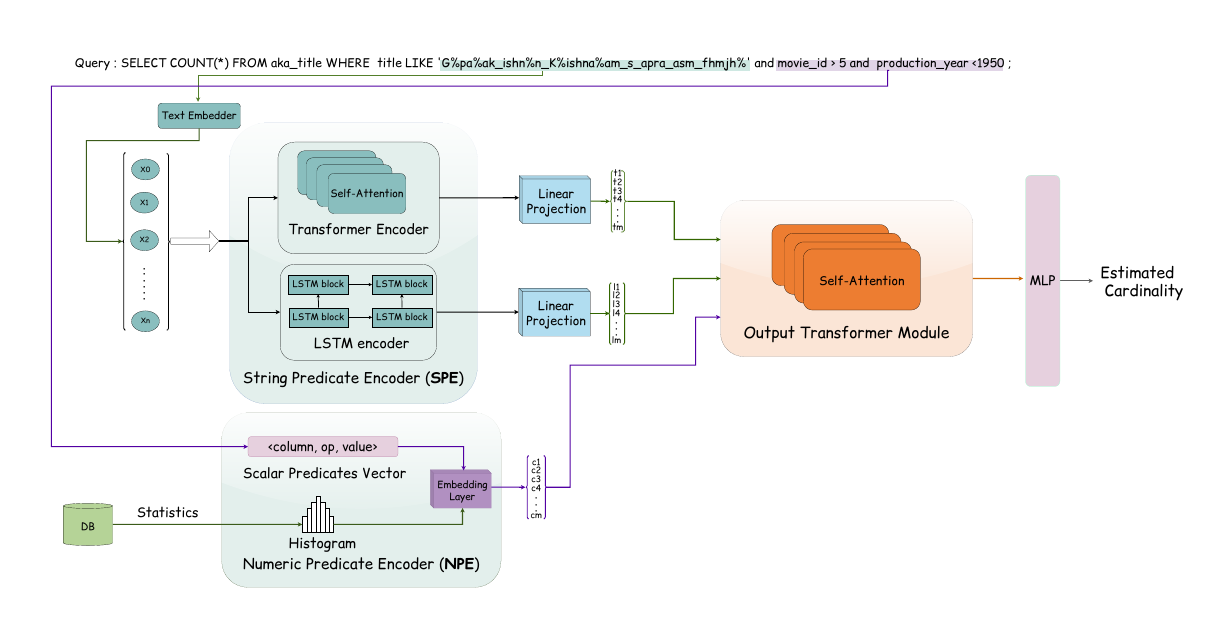}
  \caption{The architecture of MultiCard. First, the string predicate is processed by an embedding layer to generate a dense vector representation. This vector is then passed to the String Predicate Encoder block, which combines a Transformer to capture long-range textual dependencies and an LSTM to handle shorter patterns. Simultaneously, numerical predicates are transformed into representation vectors and combined with histogram data through a dedicated numerical embedder. Finally, these three feature vectors are fused using a Transformer layer to learn their cross-feature interactions, and a Multi-Layer Perceptron (MLP) outputs the final cardinality estimation.}
  \label{fig:multicard-arch}
\end{figure}

\textbf{Data-driven methods:} Techniques such as \cite{hilprecht2019deepdb}, \cite{yang2020neurocard}, \cite{zhu2020flat}, \cite{wang2021face}, \cite{wu2023factorjoin}, \cite{kim2024asm}, and \cite{zhang2025distjoin} construct models by analyzing data distributions directly from the database tables, completely independent of query workloads. While these methods generally provide high accuracy and stable estimates, they suffer from two primary limitations: they struggle to accurately capture complex data relationships in non-equi joins, and adapting them to dynamic data updates remains highly challenging.

\textbf{Query-driven methods:} Models such as\cite{kipf2018learned}, \cite{dutt2019selectivity}, \cite{sun2019end}, \cite{aytimur2025space}, \cite{ortiz2019empirical}, \cite{negi2023robust}, \cite{li2023alece}, and \cite{zhu2020flat} learn by training on a large set of real queries. Similarly, certain models like \cite{kwon2025cardinality}, \cite{kwon2022cardinality}, and \cite{aytimur2024lplm} learn from sets of strings. This makes them highly flexible for specific workloads, and they do not require heavy computational power to train. However, their accuracy degrades significantly when query results vary widely, such as in non-equi joins. Furthermore, gathering and executing a sufficient number of queries to train or update these models is highly expensive. Nevertheless, recent studies have addressed these real-world challenges of query-driven approaches; for instance, while query-driven methods traditionally struggle with the out-of-distribution (OOD) problem, \cite{salihouglu2026cardinality} successfully resolves this issue. Additionally, new work improves estimation accuracy by combining online learning with query-driven methods \cite{yi2025bigger}.

\textbf{Hybrid methods:} Approaches such as those proposed by \cite{wu2021unified}, \cite{zhang2024duet}, and \cite{liu2025unified} attempt to combine the strengths of both worlds by integrating data analysis with query experience. Similarly, \cite{zeng2024price} introduced a pre-trained hybrid model. Additionally, recent studies have explored agentic methods for this problem \cite{wehrstein2026bespoke}. While this combination yields the highest estimation accuracy, it comes at the cost of prolonged training times and substantial memory footprints. Our proposed model also falls into the hybrid category; we accept the trade-offs of increased training time and a larger model footprint because they enable us to achieve significantly superior estimation accuracy.


\definecolor{hdrblue}{RGB}{220, 230, 242}   
\definecolor{multigray}{RGB}{242, 242, 242}  

\begin{table}[htbp]
\centering
\small
\caption{Qualitative comparison of cardinality estimation paradigms across key evaluation criteria.}
\label{tab:ce_comparison}
\renewcommand{\arraystretch}{1.2}
\setlength{\tabcolsep}{4pt}

\begin{tabular}{%
    >{\raggedright\arraybackslash}p{2.8cm}
    >{\centering\arraybackslash}p{1.1cm}
    >{\centering\arraybackslash}p{1.1cm}
    >{\centering\arraybackslash}p{1.1cm}
    >{\centering\arraybackslash}p{1.1cm}
}
\toprule

\textbf{Criterion}
& \textbf{Trad.}
& \textbf{Query-Driven}
& \textbf{Data-Driven}
& \textbf{Hybrid}\\
\midrule

Estimation Accuracy
& L
& M
& M
& \textbf{H}\\

Data Dependency Awareness
& L
& L
& \textbf{H}
& \textbf{H}\\

Query Workload Awareness
& N
& \textbf{H}
& N
& \textbf{H}\\

Training Cost
& N
& M
& M-H
& H\\

Inference Latency
& L
& \textbf{L}
& H
& \textbf{L}\\

Memory
& \textbf{L}
& M-H
& M
& H\\

\bottomrule
\multicolumn{5}{l}{%
\footnotesize L/M/H denote Low/Medium/High, respectively; N denotes None.}\\
\multicolumn{5}{l}{%
\footnotesize Compound ratings (e.g., M-H) indicate variability across representative systems.}

\end{tabular}
\label{comparefeature}
\end{table}
\begin{table}[htbp]
\centering
\caption{Type Of Strings (LIKE Patterns) Supported By SPE}
\begin{tabular}{|c|c|c|}
\hline
Query Category      & Description                                    & Example Pattern      \\ \hline
Exact Query         & Matches the complete literal string            & Still Orangutans     \\ \hline
Prefix Query        & Matches strings starting with specific text    & Sti\%                \\ \hline
Suffix Query        & Matches strings ending with specific text      & \%Orangutans         \\ \hline
Substring Query     & Matches strings containing specific text       & S\%Oran\%            \\ \hline
Positional Query    & Matches strings with specific character counts & Sti\_l\_Oran\_utan\_ \\ \hline
Complex Combination & Any mixed usage of \% and \_ wildcards         & \%ill\_Or\%ta\_s     \\ \hline
\end{tabular}
\label{compare_QUEryMEthod}
\end{table}

\section{Methodology}

To estimate the cardinality of multi-attribute queries including string LIKE predicates and numerical filters, we have constructed an end-to-end model with a deep unified architecture. Because the model requires fundamentally different strategies for processing LIKE string patterns versus numerical range filters, we have designed independent components to handle each type of predicate. The proposed architecture (illustrated in Figure~\ref{fig:multicard-arch}) comprises the following modules:

\subsection{String Predicate Encoder (SPE)}

In this section, we employ a hybrid architecture combining a Long Short-Term Memory (LSTM) network and a Transformer named SPE block. The rationale behind this dual approach is to leverage the Transformer's superior capability in handling long-range LIKE patterns; while utilizing the LSTM for effectively modeling shorter sequences. By integrating both architectures, we aim to accurately detect and represent patterns across varying lengths. Initially, the input data—represented as a sequence of tokens \( (x_1, x_2, \dots, x_n) \)
is processed by a parser and tokenizer before entering the LIKE Encoder architecture.

\subsubsection*{Transformer Encoder}

This component utilizes the Self-Attention mechanism. Self-Attention allows the model to attend to the entire sequence simultaneously during the processing of each token, thereby comprehending the interrelationships between them. This feature enables the robust identification of long-range dependencies within extended sequences.

\subsubsection*{LSTM Encoder}

We use LSTM blocks, which are used for sequential data processing. In our architecture, three stacked LSTM blocks are utilized sequentially to extract features and capture temporal dynamics during a like pattern.

Each of these two components independently generates an embedding vector to represent the LIKE pattern. The output of the LSTM module corresponds to the last hidden state, extracted via the following $h_{\text{last}} = \mathrm{LSTM}(x_1, x_2, \dots, x_n)$.
Similarly, the output of the Transformer layer corresponds to the last valid timestep based on the sequence lengths. In summary, the LIKE Encoder yields two distinct embedding vectors, ensuring high accuracy in processing both short-term and long-term LIKE patterns simultaneously. Table~\ref{compare_QUEryMEthod} presents a comprehensive list of all patterns supported by the proposed model.
\subsection{Numeric Predicate Encoder (NPE)}

One of the core challenges in learning the cardinality estimation for relational databases is effectively representing filter predicates for deep learning models. Traditional approaches often handle the query's structural information and the database catalog's statistical information separately, typically through a late fusion strategy. This separation limits the model's ability to capture direct, non-linear interactions between specific operators and the underlying data distributions.

To overcome this limitation, we adopt an early fusion approach at the feature extraction stage. Here, statistical information derived from histograms serves as a deterministic anchor and is fused with the filter's structural features at the earliest possible point. The process consists of the following stages:

\subsubsection{Composite Feature Formulation}

Consider a query with $F$ filter predicates. For each filter 
$i \in \{1, \dots, F\}$, we construct a comprehensive heterogeneous feature vector 
$\mathbf{f}_i \in \mathbb{R}^{d_{\text{feat}}}$ including structur and statistical features. This vector concatenates two complementary components:

\begin{itemize}
    \item \textbf{Structural and textual features} 
    ($\mathbf{v}_{\text{struct}}^{(i)}$): These encode the raw filter information, including the column ID, the conditional operator (e.g., $=$, $<$, $>$) in one-hot encoding, and normalized filter values.
    
    \item \textbf{Database statistical features} 
    ($\mathbf{v}_{\text{hist}}^{(i)}$): These consist of selectivity related statistics extracted directly from the database catalog's one-dimensional histograms for the relevant column and range.
\end{itemize}

The final feature vector for the $i$-th filter is formed by concatenation:

\begin{equation}
\mathbf{f}_i = 
\left[
\mathbf{v}_{\text{struct}}^{(i)} \,\|\, 
\mathbf{v}_{\text{hist}}^{(i)}
\right]
\in \mathbb{R}^{d_\text{feat}}
\end{equation}

Performing this concatenation enables the neural network to learn complex, non-linear interactions between the operator/column type and the corresponding statistical estimates from the very first layers. For example, the model can learn how much to trust or adjust the catalog's selectivity estimate depending on the operator (e.g., equality vs.\ inequality) and the degree of skewness in a particular column.

\subsubsection{Filter Embedder Layer}

Because each filter feature vector $f_i$ contains a heterogeneous mixture of one-hot categorical values, normalized continuous values, and statistical probabilities, it cannot be directly processed by interactive sequence-modeling layers. Instead, each $f_i$ is passed through a dedicated MLP which maps the raw features into a dense token within a continuous embedding space.

To improve gradient stability and mitigate overfitting, the MLP incorporates Layer Normalization and Dropout. The encoding process is executed in two steps:

\begin{equation}
h_1^{(i)} =
\text{Dropout}
\left(
\text{ReLU}
\left(
\text{LayerNorm}(W_1 f_i + b_1)
\right)
\right)
\end{equation}

\begin{equation}
t_{\text{filter}}^{(i)} =
\text{Dropout}
\left(
\text{ReLU}
\left(
\text{LayerNorm}(W_2 h_1^{(i)} + b_2)
\right)
\right)
\end{equation}

where
$W_1 \in \mathbb{R}^{d_{\text{model}} \times d_{\text{feat}}}$
and
$W_2 \in \mathbb{R}^{d_{\text{model}} \times d_{\text{model}}}$
are learnable weight matrices. This process yields an enriched filter representation,
$t_{\text{filter}}^{(i)} \in \mathbb{R}^{d_{\text{model}}}$.

\subsubsection{Projection to Fusion Space}

Finally, to align the filter tokens with tokens derived from other query components (such as text patterns found in \texttt{LIKE} predicates), we apply a linear projection layer:

\begin{equation}
t_{\text{fusion}}^{(i)}
=
W_{\text{proj}} t_{\text{filter}}^{(i)}
+
b_{\text{proj}}
\in
\mathbb{R}^{d_{\text{fusion}}}
\end{equation}

where
$W_{\text{proj}}
\in
\mathbb{R}^{d_{\text{fusion}} \times d_{\text{model}}}$
and
$b_{\text{proj}}
\in
\mathbb{R}^{d_{\text{fusion}}}$
are learnable parameters.

\subsection{Output Transformer Module and Final Layer}

Once all embedding vectors from the three encoders are projected into a unified dimensional space, they are concatenated and passed as the combined feature set to the final Transformer.
The core of this stage is the Self-Attention mechanism within the Transformer. This allows the model to capture intricate dependencies and cross-component interactions between the LIKE pattern embeddings (from both LSTM and Transformer encoders) and the histogram-fused numerical filter embeddings. Unlike simple linear combinations, Self-Attention enables the model to dynamically weight the importance of each feature based on the context of the specific query, effectively synthesizing the most relevant information for the final estimation.
Following the Transformer layers, a Multi-Layer Perceptron (MLP) serves as the final regression head. The MLP acts as a non-linear regressor that processes the refined features from the Transformer to generate the final cardinality prediction. The MLP is crucial for learning complex, high-order combinations of features that cannot be captured by linear projections alone.

\subsection{Training and Loss Function}
To optimize training efficiency and avoid redundant data processing, we utilize the data serialization. The raw dataset is pre-processed and stored as PKL files before being loaded for training and evaluation.

\subsubsection*{End-to-end Optimization}

Our architecture is trained in an end-to-end fashion. Consequently, the loss signals from the final output propagate back through all components. 
For loss function, we employ the Mean Squared Error (MSE) on the logarithmic scale as our objective function. Training on the log-transformed cardinality is essential because cardinality values can span several orders of magnitude. The loss function of model is in Log-MSE format defined as:
\begin{equation}
L_{\text{model}} = \frac{1}{N} \sum_{i=1}^{N} (\log(\hat{C}_i) - \log(C_i))^2
\end{equation}
Where \( \hat{C}_i \) is the predicted cardinality and \( C_i \) is the ground truth.

\section{Experiments}

\subsection{Experimental Setup}
The implementation was carried out using Python 3.10.12 and PostgreSQL 17.5. All experiments were conducted on NVIDIA GeForce RTX 3090 Ti with 24GB of VRAM.
We employed the Optuna framework for hyperparameter optimization. The tuning process involved searching across various architectural and training hyperparameters, including the embedding dimensions, the number of layers and hidden units in the LSTM, Transformer parameters, dropout rates, and fusion layer dimensions, as well as the learning rate and weight decay. In each trial, the model was trained on the training set, and its performance was evaluated using the mean $Q\text{-error}$ on the validation set. To mitigate overfitting and reduce computational overhead, we integrated an Early Stopping mechanism alongside Optuna's Median Pruner. The optimal model configuration was selected and saved based on the minimum mean $Q\text{-error}$. The optimization process was conducted for 50 trials for the SPE model and 25 trials for the MultiCard model.

\subsection{Datasets}

We utilized several datasets in our experiments (see Table~\ref{tab:dataset-stats}). From each dataset, we selected a single table containing suitable attributes. These datasets and their corresponding attributes are described as follows:
\begin{table}[htbp]
\caption{Dataset statistics}
\centering
\resizebox{\linewidth}{!}{%
\begin{tabular}{|l|llll|lll|}
\hline
         & \multicolumn{4}{c|}{String Set in a Table}                                                      & \multicolumn{3}{c|}{Integer Set in a Table}                                      \\ \hline
Dataset  & \multicolumn{1}{l|}{$|\Sigma|$}  & \multicolumn{1}{l|}{|SD|}      & \multicolumn{1}{l|}{AVG Length} &MAX Length & \multicolumn{1}{l|}{MIN Value} & \multicolumn{1}{l|}{MAX Value} & \#Columns\\ \hline
DBLP     & \multicolumn{1}{l|}{73}   & \multicolumn{1}{l|}{2,365,100} & \multicolumn{1}{l|}{14.2}  & 62    & \multicolumn{1}{l|}{1}         & \multicolumn{1}{l|}{2,365,100} & 2              \\ \hline
WIKI     & \multicolumn{1}{l|}{3735} & \multicolumn{1}{l|}{1,031,930} & \multicolumn{1}{l|}{98.4}  & 300   & \multicolumn{1}{l|}{\_\_}      & \multicolumn{1}{l|}{\_\_}      & \_\_           \\ \hline
IMDB     & \multicolumn{1}{l|}{81}   & \multicolumn{1}{l|}{1,000,000} & \multicolumn{1}{l|}{19.2}  & 278   & \multicolumn{1}{l|}{0}         & \multicolumn{1}{l|}{2,525,672} & 7              \\ \hline
TPCH     & \multicolumn{1}{l|}{26}   & \multicolumn{1}{l|}{200,000}   & \multicolumn{1}{l|}{32.75} & 50    & \multicolumn{1}{l|}{1}         & \multicolumn{1}{l|}{200,000}   & 3              \\ \hline
Ergastf1 & \multicolumn{1}{l|}{30}   & \multicolumn{1}{l|}{840}       & \multicolumn{1}{l|}{7.1}   & 18    & \multicolumn{1}{l|}{1}         & \multicolumn{1}{l|}{841}       & 2              \\ \hline
\end{tabular}%
}
\label{tab:dataset-stats}
\end{table}

\begin{itemize}
    \item \textbf{IMDB:} We used the \texttt{aka\_title} table from this dataset. The attribute selected for string data is \texttt{title}, which represents movie titles. The integer attributes selected are \texttt{movie\_id}, \texttt{production\_year}, and \texttt{id} \cite{imdb}.
    
    \item \textbf{TPCH:} From this dataset, we utilized the \texttt{part} table. The string attribute is \texttt{p\_name}, and the integer attributes are \texttt{p\_size} and \texttt{p\_partkey} \cite{tpch}.
    
    \item \textbf{DBLP:} We worked with the \texttt{author\_name} table from the DBLP dataset. The integer attributes used are \texttt{freq} and \texttt{id} \cite{dblp}.
    
    \item \textbf{ErgastF1:} We used the \texttt{drivers} table from this dataset. The string attribute selected is \texttt{driverRef}, while the integer attributes are \texttt{driverId} and \texttt{number} \cite{ergastf1}.
    
    \item \textbf{WIKI:} We utilized Wikipedia articles from the DocRED dataset. This dataset contains only string attributes; therefore, it was used exclusively to evaluate the string component of our approach \cite{yao2019docred}.
\end{itemize}

\subsection{Query Set Generation}

To evaluate the performance of our unified cardinality estimation model and train it on appropriate queries, we developed a multi-stage data generation pipeline called the Unified Pattern-Condition Interleaver (UPCI). The primary goal of UPCI is to construct a balanced dataset that prevents model bias toward specific string lengths or predicate types. UPCI consists of three main phases: string pattern generation, numerical predicate construction, and query synthesis.

For each dataset, we adopt a standardized naming convention to refer to the query sets produced by UPCI. For DBLP, we denote the string query set as DBLP-S (DBLP String Query Set), with an average string length of 17.99, a maximum string length of 34, and 291,353 unique string queries. Similarly, we denote the mixed query set as DBLP-M (DBLP Mixed Query Set), consisting of 400,000 queries that combine \texttt{LIKE} patterns with 1-to-3 numerical predicates. We follow the same naming convention for the remaining datasets: TPCH-S, with an average string length of 19.9, a maximum string length of 47, and 625,450 unique string queries, and TPCH-M, consisting of 300,000 mixed queries with up to 2 numerical predicates. For ErgastF1, ErgastF1-S has an average string length of 5.02, a maximum string length of 10, and 41,470 unique string queries, while ErgastF1-M consists of 150,000 mixed queries with up to 2 numerical predicates.

We also construct WIKI-S, with an average string length of 36.22, a maximum string length of 70, and 710,835 unique string queries, to evaluate the string-only component of our approach on a large-vocabulary dataset with long string lengths. However, since WIKI does not contain any numerical attributes, it cannot be used to construct a corresponding mixed query set. To nonetheless assess mixed-query performance on a dataset of comparable complexity, we instead construct IMDB-M, consisting of 300,000 queries that combine \texttt{LIKE} patterns with up to 3 numerical predicates.
\subsubsection{LIKE Pattern Generation}
The string patterns are generated as Algorithm~\ref{like-gen} to ensure a uniform distribution across various lengths, which is critical for the model’s robustness on both short and long sequences. 
\begin{itemize}
    \item \textbf{Short Patterns (Length < 10):} We perform random sampling from all existing rows in the PostgreSQL database. This ensures that the dataset contains a representative number of short-string samples to satisfy the required frequency for each length.
    \item \textbf{Long Patterns (Length $\geq$ 10):} To handle longer sequences, we select a base row at random and inject wildcards (\texttt{\%} or \texttt{\_}) at random positions until the target length is achieved. This approach ensures comprehensive coverage of all \texttt{LIKE} types, including \textbf{prefixes}, \textbf{suffixes}, and \textbf{complex internal patterns}.
\end{itemize}

\subsubsection{Numerical Predicate Construction}

Parallel to string generation, we construct numerical filters to capture diverse range and equality constraints as formalized in Algorithm~\ref{num-gen}.

\begin{enumerate}
    \item \textbf{Atomic Filters:} For each numerical column, we generate a set of primary filters by combining column names with various comparison operators (e.g., $<$, $\leq$, $>$, $\geq$, $=$) and values sampled from the column's domain.
    \item \textbf{Cardinality Verification:} The true record count for each atomic filter is calculated using SQL \texttt{WHERE} clauses in the PostgreSQL environment.
    \item \textbf{Composite Filters:} To simulate complex real-world queries, we combine multiple atomic filters using the \texttt{AND} operator. These combinations include 2-condition and 3-condition predicates, allowing the model to learn cross-column correlations.
\end{enumerate}

\subsubsection{Hybrid Query Synthesis and Ground Truth Labeling}

In the final stage, we synthesize the previously generated components to create realistic, complex queries. This process is designed to bridge the gap between isolated predicates and real-world database workloads:

\begin{itemize}
    \item \textbf{Mixed Query Construction:} We combine the generated \texttt{LIKE} patterns and numerical filters using a randomized round-robin strategy. By merging these two distinct types of predicates, we produce ``Mixed Queries'' that closely resemble the multi-dimensional filters found in production environments.
    \item \textbf{Ground Truth Computation:} To provide the model with accurate labels for training, each synthesized query is executed as a live SQL command against the PostgreSQL engine. We compute the exact record count (\textbf{Ground Truth}) for every combination.
\end{itemize}

 \SetKwProg{Procedure}{Procedure}{}{}
\SetKwComment{Comment}{// }{}
\SetNlSty{textbf}{}{.} 
\SetAlgoVlined 

\begin{algorithm}\small
\DontPrintSemicolon
\caption{Generate LIKE Patterns}
\label{like-gen}
\Procedure{GenerateLikeWorkload($T, S$)}{
     \KwIn{Table $T$, Column $S$}
     \KwOut{$P_{like}$}
     \BlankLine

     $P_{like} \leftarrow \emptyset; \quad seen \leftarrow \emptyset$\;
     $R \leftarrow \text{LoadAndFilterStrings}(T, S)$\;
     $L \leftarrow \text{GetFeasibleLengths}(R)$\;
    
     \For{each length $\ell \in L$}{
         \While{$|P_{like}^{(\ell)}| < Max\_Pattern\_l$}{
             $r \leftarrow \text{Sample}(R, |r| \ge \ell)$\;
            
             \Comment*[l]{Generate candidates based on length}
             \eIf{$\ell < 10$}{
                 $Cand \leftarrow \{ \text{MakeShortPattern}(r, \ell) \}$\;
             }{
                 $Cand \leftarrow \text{MakeLongPatterns}(r, \ell)$\;
             }
            
             \Comment*[l]{Filter and count in parallel}
             $ValidCand \leftarrow \{ p \in Cand \mid p \notin seen \land \text{Valid}(p) \}$\;
             $counts \leftarrow \text{ParallelCount}(T, S, ValidCand)$\;
            
             \For{each $(p, card) \in counts$}{
                 \If{$card \ge \text{cfg.min\_card}$}{
                     $P_{like} \leftarrow P_{like} \cup \{(p, card)\}$\;
                     $seen \leftarrow seen \cup \{p\}$\;
                 }
             }
         }
     }
     \Return $P_{like}$\;
}
\end{algorithm}

\begin{algorithm}
\small
\DontPrintSemicolon
\caption{Generate Numerical Predicates}
\label{num-gen}
\Procedure{GenerateNumericalWorkload($T, C, \text{cfg}$)}{
     \KwIn{Table $T$, Columns $C = \{c_1..c_k\}$, cfg}
     \KwOut{$F_{num}$}
     \BlankLine

     $F_{num} \leftarrow \emptyset$\;
     $per\_col\_target \leftarrow \text{cfg.N1} / |C|$\;

     \Comment*[l]{Phase 1: Atomic Predicates (1-condition)}
     \For{each column $c \in C$}{
         $V \leftarrow \text{SampleDistinctValues}(T, c, \text{cfg})$\;
         \For{each $v \in V$ and $op \in \{>, \ge, <, \le, =\}$}{
             $pred \leftarrow (c \ op \ v)$\;
             $card \leftarrow \text{CountWhere}(T, pred)$\;
            
             \If{$card \ge \text{cfg.min\_support}$}{
                 $F_{num} \leftarrow F_{num} \cup \{(pred, card, \text{cols}=\{c\})\}$\;
             }
            \If{$|F_{num}^{(c)}| \ge per\_col\_target$}{\textbf{break}}
         }
     }

     \Comment*[l]{Phase 2: Multi-condition (Cross-column)}
     \For{$k \in \{2, 3\}$}{
         $F_k \leftarrow \text{GenerateCombos}(F_{num}, k, \text{target}=\text{cfg.N}_k)$\;
         $F_{num} \leftarrow F_{num} \cup F_k$\;
     }

     \Return $F_{num}$\;
}
\end{algorithm}

The resulting samples, containing the combined predicates and their corresponding ground truth labels, are then shuffled and split into training, validation, and testing sets, and finally serialized for optimized model consumption.

\section{Baseline Methods}

To our knowledge, no existing learned cardinality estimator jointly handles LIKE string predicates and numerical range filters. Models such as CLIQUE, LPLM and DREAM are restricted to string only workloads, while data-driven and hybrid methods such as DeepDB, NeuroCard, and PRICE target numerical and join predicates exclusively. Consequently, PostgreSQL's rule-based estimator is the only system against which MultiCard can be fairly evaluated on multi-attribute queries.

Various models have been proposed in the domain of cardinality estimation. While some approaches specifically target cardinality estimation for LIKE queries, others focus exclusively on numerical filters. In this section, we briefly describe the baseline models selected for comparison in this study. Although extensive research exists in this field, we have examined three state-of-the-art methods as compared in Table \ref{tbl:relatedcompare}. As demonstrated in their respective publications, these models have achieved superior performance across all evaluated datasets.

\begin{table}[htbp]
\centering
\caption{Classification of the methods considered in this paper.}
\label{tab:method-classification}
\begin{tabular}{lccc}
\toprule
Method & Query-driven & Data-driven & Hybrid \\
\midrule
CLIQUE    & \ding{51} &           &           \\
DREAM     & \ding{51} &           &           \\
LPLM      & \ding{51} &           &            \\
SPE (ours)& \ding{51} &           &           \\
PostgreSQL&           & \ding{51} &           \\
MultiCard (ours) &    &           & \ding{51} \\
\bottomrule
\end{tabular}
\label{tbl:relatedcompare}
\end{table}

\subsection{CLIQUE}
CLIQUE \cite{kwon2025cardinality} utilizes an $n$-gram table to establish an upper bound for cardinality. Subsequently, it refines the estimate generated by an LSTM-based model using this upper bound to achieve a more precise cardinality prediction.

\subsection{LPLM}
LPLM~\cite{aytimur2024lplm} starts from the observation that many pioneering
methods, are primarily designed for patterns in which wildcards appear only at
the beginning or the end of the string, and perform poorly on patterns where a
wildcard is placed between other characters (e.g., \texttt{a\%b\_c}). To
overcome this limitation, LPLM introduces a neural language model built upon a
new language for representing \texttt{LIKE} patterns together with a
novel probability distribution function; this function is designed to
capture the full semantics of a general \texttt{LIKE} pattern, regardless of the
positions of the wildcards within it.

\subsection{DREAM}
DREAM \cite{kwon2022cardinality} introduces a method for estimating the cardinality of approximate substring queries using an LSTM-based deep learning architecture.
The DREAM model processes the query string and an error threshold $(\epsilon)$ sequentially across characters, simultaneously learning the result counts for both the query itself and all its prefixes. Furthermore, it employs a novel training strategy known as \emph{packed learning} to accelerate the process and enhance accuracy.

\subsection{PostgreSQL}
The PostgreSQL cardinality estimator is a purely statistical, rule-based component that relies on data collected via the ANALYZE command—such as histograms and Most Common Values (MCVs)—to approximate row counts for query planning. While it calculates simple predicates effectively, text-based conditions like LIKE rely on simplistic assumptions and require extensions like pg\_trgm for accuracy. Furthermore, multiple concurrent conditions suffer from an often-flawed independence assumption unless extended statistics are applied. Despite its lack of machine learning and strict dependence on statistical quality, PostgreSQL remains the only evaluated method capable of handling multi-attribute queries.
\subsection{Sampling}
To the best of our knowledge, no existing learned cardinality estimator jointly supports \texttt{LIKE} predicates and numerical range predicates in the same query. We therefore adopt uniform sampling with two sampling rates, $0.1\%$ and $1\%$, as our main baseline, since it supports both predicate types natively and, by evaluating predicates on real tuples, preserves the correlations between string and numerical attributes without any independence assumption. Concretely, for a sampling rate $p$ we materialize a sample of the base table once, where each tuple is independently included with probability $p/100$. At estimation time, each test query is executed unchanged on the sample, and its count $c$ is scaled by the inverse sampling rate, giving the unbiased estimate $\hat{c} = c \cdot 100/p$. The main limitation of sampling appears for long \texttt{LIKE} patterns: as a pattern becomes longer and more specific, its selectivity drops sharply, so few or no sampled tuples match it. The estimate then collapses to zero or is dominated by the variance of a handful of matching tuples, which produces large q-errors, especially at the lower $0.1\%$ rate. In addition, every estimate requires a scan of the sample, so inference time grows linearly with the sample size.

\section{Evaluation}
To assess the accuracy of various methods in estimating query result counts, we use the \emph{Q-error} metric. This metric quantifies the relative multiplicative difference between the actual cardinality and the estimated cardinality, irrespective of the direction of the error (i.e., whether it is an over-estimation or under-estimation).

Let $C$ denote the actual cardinality and $\hat{C}$ denote the estimated cardinality. The Q-error is defined as follows:
\begin{equation}
\text{Q-error} = \max \left( \frac{C}{\hat{C}}, \frac{\hat{C}}{C} \right)
\end{equation}

The optimal value for this metric is 1, which is achieved when the estimated value exactly matches the ground truth. As the deviation between the two values increases, the Q-error grows accordingly. Theoretically, this metric has no upper bound.
In this study, we measured inference times to assess the model's efficiency. Test latency (inference time) and evaluation metrics represent two complementary facets of model performance analysis. Evaluation metrics are utilized to gauge the accuracy and quality of the model's output, determining the extent to which the generated results align with the actual values. Conversely, test time indicates the duration required for the model to process new data and generate predictions following the training phase.
We evaluate the model size metric to assess the structural complexity and resource demands of each approach. This metric encompasses the total number of learnable parameters and the storage footprint required by the model. It serves as a proxy for understanding the financial and computational overhead of training, as larger architectures necessitate more expensive hardware and longer execution times.
\subsection{Performance Evaluation}
According to Table~\ref{SPE:comparison}, SPE consistently provides accurate estimations, frequently outperforming others across datasets. SPE demonstrates superior performance in average and high-percentile Q-errors over most existing approaches, signifying its effectiveness in representing intricate data patterns and yielding more dependable cardinality estimates. Beyond enhanced accuracy, SPE provides a desirable equilibrium between the quality of estimation and computational speed. While certain traditional methods like PostgreSQL offer quicker inference, they come with significantly larger estimation errors. SPE performs as well as learning-based baselines in terms of inference time, but it provides superior estimation results. SPE’s marginally larger number of parameters is warranted by its enhanced representation and considerably reduced estimation errors, indicating efficient use of model complexity. Although SPE contains more parameters than CLIQUE, LPLM, and DREAM in most datasets, it consistently achieves lower Q-error values, demonstrating that the increased model capacity is effectively utilized rather than introducing unnecessary overhead. \\
Across all datasets shown in Table~\ref{MultiCard:comparison}, MultiCard surpasses PostgreSQL by a considerable margin, showing substantially reduced average and percentile Q-errors. Traditional estimators struggle with complex data distributions, but MultiCard’s improvements demonstrate its ability to model attribute correlations effectively. Despite a minor increase in inference time, MultiCard delivers notably superior and more dependable estimates, all contained within a single model, underscoring its value for practical cardinality estimation.
\begin{table}[htbp]
\centering
\small
\setlength{\tabcolsep}{4pt}
\caption{Comparative analysis of SPE and existing baselines on LIKE pattern (single-attribute) queries. (The best results are highlighted in dark green, while the second-best results are highlighted in light green)}
\begin{tabular}{|c|c|ccccc|c|c|}
\hline
                           &                           & \multicolumn{5}{c|}{Q-ERROR}                                                                                                                                &                                  &                                \\ \cline{3-7}
\multirow{-2}{*}{Datasets} & 
\multirow{-2}{*}{Methods} & 
50th & 
95th & 
99th & 
AVG & 
MAX & 
\multirow{-2}{*}{\makecell{Inference Time\\(ms)}} & 
\multirow{-2}{*}{\#Params} \\ \hline
                           & CLIQUE                    & \cellcolor[HTML]{EFF4ED}1.27 & \cellcolor[HTML]{EFF4ED}3.58 & \cellcolor[HTML]{EFF4ED}12.1 & 2.58                         & 9895.44                         & 1.72                             & \cellcolor[HTML]{C6E0C2}2.42M  \\
                           & DREAM                     & 1.35                         & 4.27                         & 15.19                        & \cellcolor[HTML]{C6E0C2}2.4  & \cellcolor[HTML]{C6E0C2}1487.82 & 1.61                             & \cellcolor[HTML]{EFF4ED}2.43M  \\
                           & PostgreSQL                  & 14.33                        & 43                           & 145.96                       & 24.19                        & 51591.35                        & \cellcolor[HTML]{C6E0C2}0.08     & ---                            \\
\multirow{-4}{*}{WIKI-S}     & LPLM                      & 2                            & 18.24                        & 27                           & 5.2                          & 10876.6                         & 4.62                             & 2.44M                          \\
                           & SPE (ours)                & \cellcolor[HTML]{C6E0C2}1.25 & \cellcolor[HTML]{C6E0C2}3.54 & \cellcolor[HTML]{C6E0C2}12   & \cellcolor[HTML]{EFF4ED}2.52 & \cellcolor[HTML]{EFF4ED}9895    & \cellcolor[HTML]{EFF4ED}1.53     & 2.84M                          \\ \hline
                           & CLIQUE                    & \cellcolor[HTML]{EFF4ED}1.13 & \cellcolor[HTML]{EFF4ED}3.48 & \cellcolor[HTML]{EFF4ED}8.02 & \cellcolor[HTML]{EFF4ED}1.6  & \cellcolor[HTML]{C6E0C2}117.87  & 0.86                             & \cellcolor[HTML]{C6E0C2}0.58M  \\
                           & DREAM                     & 1.21                         & 5.56                         & 16.66                        & 3.96                         & 14349.3                         & \cellcolor[HTML]{EFF4ED}0.83     & \cellcolor[HTML]{EFF4ED}0.59M  \\
                           & PostgreSQL                  & 118.5                        & 118.5                        & 118.5                        & 81.04                        & 11945                           & \cellcolor[HTML]{C6E0C2}0.07     & ---                            \\
\multirow{-4}{*}{DBLP-S}     & LPLM                      & 2.0                          & 47.0                         & 256.38                       & 15.34                        & 4885                            & 1.27                             & 0.61M                          \\
                           & SPE (ours)                & \cellcolor[HTML]{C6E0C2}1.09 & \cellcolor[HTML]{C6E0C2}3.42 & \cellcolor[HTML]{C6E0C2}7.62 & \cellcolor[HTML]{C6E0C2}1.59 & \cellcolor[HTML]{EFF4ED}146.71  & 0.97                             & 0.93M                          \\ \hline
                           & CLIQUE                    & \cellcolor[HTML]{EFF4ED}1.01 & 2.13                         & \cellcolor[HTML]{EFF4ED}4.36 & \cellcolor[HTML]{EFF4ED}1.21 & 79.87                           & 0.32                             & \cellcolor[HTML]{C6E0C2}0.56M  \\
                           & DREAM                     & 1.02                         & \cellcolor[HTML]{EFF4ED}2.12 & 4.39                         & 1.22                         & \cellcolor[HTML]{EFF4ED}45.26   & 0.3                              & \cellcolor[HTML]{EFF4ED}0.57M  \\
                           & PostgreSQL                  & 2.18                         & 86.85                        & 181.46                       & 16.99                        & 876.3                           & \cellcolor[HTML]{C6E0C2}0.05     & ---                            \\
\multirow{-4}{*}{TPCH-S}     & LPLM                      & \cellcolor[HTML]{C6E0C2}1    & 7.43                         & 22                           & 2.46                         & 487                             & \cellcolor[HTML]{EFF4ED}0.42     & 0.61M                          \\
                           & SPE (ours)                & \cellcolor[HTML]{C6E0C2}1    & \cellcolor[HTML]{C6E0C2}1.87 & \cellcolor[HTML]{C6E0C2}3.1  & \cellcolor[HTML]{C6E0C2}1.15 & \cellcolor[HTML]{C6E0C2}29.09   & 0.32                             & 0.82M                          \\ \hline
                           & CLIQUE                    & \cellcolor[HTML]{EFF4ED}1.12 & \cellcolor[HTML]{C6E0C2}1.84 & \cellcolor[HTML]{C6E0C2}2.49 & \cellcolor[HTML]{EFF4ED}1.23 & \cellcolor[HTML]{C6E0C2}5.35    & 0.27                             & \cellcolor[HTML]{C6E0C2}0.27M  \\
                           & DREAM                     & 1.13                         & 1.94                         & \cellcolor[HTML]{C6E0C2}2.57 & 1.26                         & 11.85                           & 0.26                             & \cellcolor[HTML]{EFF4ED}0.28M \\
                           & PostgreSQL                  & 2                            & 5                            & 9                            & 2.77                         & 33                              & \cellcolor[HTML]{C6E0C2}0.03     & ---                            \\
\multirow{-4}{*}{ERGASTF1-S} & LPLM                      & 2.0                          & 5                            & 11.4                         & 2.26                         & 30                              & 0.31                             & 0.28M                          \\
                           & SPE (ours)                & \cellcolor[HTML]{C6E0C2}1.04 & \cellcolor[HTML]{EFF4ED}1.9  & 2.62                         & \cellcolor[HTML]{C6E0C2}1.2  & \cellcolor[HTML]{EFF4ED}5.46    & \cellcolor[HTML]{EFF4ED}0.19     & 0.45M                          \\ \hline
\end{tabular}
\label{SPE:comparison}
\end{table}

\begin{table}[htbp]
\centering
\caption{Comparative analysis of MultiCard and existing baselines on LIKE pattern and numerical predicate (multi-attribute) queries. (The best results are highlighted in dark green, while the second-best results are highlighted in light green.)}
\label{MultiCard:comparison}
\small
\resizebox{\linewidth}{!}{%
\begin{tabular}{|c|c|ccccc|c|c|}
\hline
                             &                           & \multicolumn{5}{c|}{Q-ERROR}                                                                                                                                    &                                                                                 &                            \\ \cline{3-7}
\multirow{-2}{*}{Datasets}   & \multirow{-2}{*}{Methods} & 50th                         & 95th                          & 99th                           & AVG                           & MAX                             & \multirow{-2}{*}{\begin{tabular}[c]{@{}c@{}}Inference Time\\ (ms)\end{tabular}} & \multirow{-2}{*}{\#Params} \\ \hline
                             & Sampling 0.1\%            & 3                            & 214.00                        & 853.00                         & 44.96                         & 4791.00                         & \cellcolor[HTML]{D0E5CB}0.21                                                    & ---                        \\
                             & Sampling 1\%              & \cellcolor[HTML]{EFF4ED}2.00 & 45.00                         & \cellcolor[HTML]{EFF4ED}108.00 & \cellcolor[HTML]{EFF4ED}9.43  & \cellcolor[HTML]{EFF4ED}524.00  & \cellcolor[HTML]{EFF4ED}0.68                                                    & ---                        \\
                             & PostgreSQL                & 3                            & \cellcolor[HTML]{EFF4ED}35    & 153.7                          & 10.51                         & 1838.5                          & \cellcolor[HTML]{FFFFFF}0.37                                                    & ---                        \\
\multirow{-4}{*}{IMDB-M}     & MultiCard                 & \cellcolor[HTML]{C6E0C2}1.03 & \cellcolor[HTML]{C6E0C2}1.64  & \cellcolor[HTML]{C6E0C2}2.65   & \cellcolor[HTML]{C6E0C2}1.15  & \cellcolor[HTML]{C6E0C2}67.7    & 4.47                                                                            & 1.62M                      \\ \hline
                             & Sampling 0.1\%            & 6.00                         & 406.00                        & 1267.00                        & 79.07                         & 8500.00                         & \cellcolor[HTML]{D0E5CB}0.56                                                    & ---                        \\
                             & Sampling 1\%              & 3.0                          & \cellcolor[HTML]{EFF4ED}53.00 & \cellcolor[HTML]{EFF4ED}151.00 & \cellcolor[HTML]{EFF4ED}12.69 & 8366.66                         & \cellcolor[HTML]{EFF4ED}2.3                                                     & ---                        \\
                             & PostgreSQL                & 6.4                          & 74.33                         & 198.93                         & 21.38                         & \cellcolor[HTML]{EFF4ED}2681.43 & \cellcolor[HTML]{FFFFFF}0.04                                                    & ---                        \\
\multirow{-4}{*}{DBLP-M}     & MultiCard                 & \cellcolor[HTML]{C6E0C2}1.16 & \cellcolor[HTML]{C6E0C2}1.91  & \cellcolor[HTML]{C6E0C2}3.44   & \cellcolor[HTML]{C6E0C2}1.35  & \cellcolor[HTML]{C6E0C2}354.4   & 4.09                                                                            & 1.61M                      \\ \hline
                             & Sampling 0.1\%            & \cellcolor[HTML]{D0E5CB}1.00 & 259.00                        & 985.00                         & 47.95                         & 5474.00                         & \cellcolor[HTML]{D0E5CB}0.19                                                    & ---                        \\
                             & Sampling 1\%              & \cellcolor[HTML]{EFF4ED}1.01 & \cellcolor[HTML]{EFF4ED}33.00 & \cellcolor[HTML]{EFF4ED}110.00 & \cellcolor[HTML]{EFF4ED}6.91  & \cellcolor[HTML]{EFF4ED}483.00  & \cellcolor[HTML]{EFF4ED}0.401                                                   & ---                        \\
                             & PostgreSQL                & 5                            & 44.01                         & 185.15                         & 12.21                         & 1189                            & \cellcolor[HTML]{FFFFFF}0.04                                                    & ---                        \\
\multirow{-4}{*}{TPCH-M}     & MultiCard                 & \cellcolor[HTML]{FFFFFF}1.04 & \cellcolor[HTML]{C6E0C2}2.18  & \cellcolor[HTML]{C6E0C2}3.8    & \cellcolor[HTML]{C6E0C2}1.28  & \cellcolor[HTML]{C6E0C2}80.9    & 4.55                                                                            & 1.05M                      \\ \hline
                             & Sampling 0.1\%            & 4.0                          & 27.00                         & 78.00                          & 8.82                          & 805.00                          & \cellcolor[HTML]{D0E5CB}0.14                                                    & ---                        \\
                             & Sampling 1\%              & 4.0                          & 25.00                         & 39.00                          & 7.28                          & 170                             & \cellcolor[HTML]{EFF4ED}0.16                                                    & ---                        \\
                             & PostgreSQL                & \cellcolor[HTML]{EFF4ED}2.2  & \cellcolor[HTML]{EFF4ED}6     & \cellcolor[HTML]{EFF4ED}9      & \cellcolor[HTML]{EFF4ED}2.58  & \cellcolor[HTML]{EFF4ED}24      & \cellcolor[HTML]{FFFFFF}0.03                                                    & ---                        \\
\multirow{-4}{*}{ERGASTF1-M} & MultiCard                 & \cellcolor[HTML]{C6E0C2}1.01 & \cellcolor[HTML]{C6E0C2}1.3   & \cellcolor[HTML]{C6E0C2}1.44   & \cellcolor[HTML]{C6E0C2}1.06  & \cellcolor[HTML]{C6E0C2}2.6     & 4.04                                                                            & 0.93M                      \\ \hline
\end{tabular}%
}
\end{table}

\subsection{Effect of Training Data Size on Q-error}
To understand how training data quantity influences $Q\text{-error}$, we experimented with different training set sizes. The model was trained on datasets made up of 10\%, 50\%, 80\%, and the full original training data. The corresponding results across different datasets are presented in Table~\ref{training-ratio}. When only 10\% of the training data is available, the model still achieves reasonable estimation accuracy, demonstrating the capability of MultiCard to learn meaningful patterns even with limited training samples. Although the performance gain gradually diminishes when increasing the training ratio beyond 80\%, the full training set generally achieves the best or near-best results across all datasets.
\begin{table}[htbp]
\centering
\caption{Impact of training data volume on Q-error of MultiCard across datasets.}
\setlength{\tabcolsep}{3pt}
{\small%
\begin{tabular}{|c|c|ccccc|}
\hline
                                                 &                                  & \multicolumn{5}{c|}{Q-ERROR}                                                                                                                              \\ \cline{3-7}
\multirow{-2}{*}{Datasets} & \multirow{-2}{*}{Training Ratio} & 50th                         & 95th                         & 99th                         & AVG                          & MAX                           \\ \hline
                                                 & 10\%                             & 1.16                         & 2.38                         & 4.43                         & 1.42                         & 274.9                         \\
                                                 & 50\%                             & \cellcolor[HTML]{EFF4ED}1.05 & 1.78                         & 2.94                         & 1.2                          & 505.3                         \\
                                                 & 80\%                             & 1.06                         & \cellcolor[HTML]{EFF4ED}1.72 & \cellcolor[HTML]{EFF4ED}2.66 & \cellcolor[HTML]{EFF4ED}1.18 & \cellcolor[HTML]{C6E0C2}41.5  \\
\multirow{-4}{*}{IMDB-M}                           & 100\%                            & \cellcolor[HTML]{C6E0C2}1.03 & \cellcolor[HTML]{C6E0C2}1.64 & \cellcolor[HTML]{C6E0C2}2.65 & \cellcolor[HTML]{C6E0C2}1.15 & \cellcolor[HTML]{EFF4ED}67.7  \\ \hline
                                                 & 10\%                             & 1.65                         & 6.13                         & 17.1                         & 2.97                         & 1601.1                        \\
                                                 & 50\%                             & 1.33                         & 3.19                         & 7.41                         & 1.83                         & \cellcolor[HTML]{EFF4ED}600.9 \\
                                                 & 80\%                             & \cellcolor[HTML]{EFF4ED}1.28 & \cellcolor[HTML]{EFF4ED}2.45 & \cellcolor[HTML]{EFF4ED}4.81 & \cellcolor[HTML]{EFF4ED}1.6  & 1700.9                        \\
\multirow{-4}{*}{DBLP-M}                           & 100\%                            & \cellcolor[HTML]{C6E0C2}1.16 & \cellcolor[HTML]{C6E0C2}1.91 & \cellcolor[HTML]{C6E0C2}3.44 & \cellcolor[HTML]{C6E0C2}1.35 & \cellcolor[HTML]{C6E0C2}354.4 \\ \hline
                                                 & 10\%                             & 1.09                         & 4.01                         & 9.54                         & 1.72                         & 162.4                         \\
                                                 & 50\%                             & \cellcolor[HTML]{EFF4ED}1.05 & 2.74                         & 5.54                         & 1.4                          & 153                           \\
                                                 & 80\%                             & \cellcolor[HTML]{C6E0C2}1.04 & \cellcolor[HTML]{EFF4ED}2.42 & \cellcolor[HTML]{EFF4ED}4.67 & \cellcolor[HTML]{EFF4ED}1.32 & \cellcolor[HTML]{C6E0C2}{75}                   \\
\multirow{-4}{*}{TPCH-M}                           & 100\%                            & \cellcolor[HTML]{C6E0C2}1.04 & \cellcolor[HTML]{C6E0C2}2.18 & \cellcolor[HTML]{C6E0C2}3.8  & \cellcolor[HTML]{C6E0C2}1.28 & \cellcolor[HTML]{EFF4ED}80.9  \\ \hline
                                                 & 10\%                             & 1.13                         & 1.75                         & 2.29                         & 1.22                         & 12                            \\
                                                 & 50\%                             & 1.04                         & 1.36                         & 1.66                         & 1.1                          & 3.7                           \\
                                                 & 80\%                             & \cellcolor[HTML]{EFF4ED}1.03 & \cellcolor[HTML]{EFF4ED}1.32 & \cellcolor[HTML]{EFF4ED}1.52 & \cellcolor[HTML]{EFF4ED}1.07 & \cellcolor[HTML]{EFF4ED}3.5   \\
\multirow{-4}{*}{ERGASTF1-M}                       & 100\%                            & \cellcolor[HTML]{C6E0C2}1.01 & \cellcolor[HTML]{C6E0C2}1.3  & \cellcolor[HTML]{C6E0C2}1.44 & \cellcolor[HTML]{C6E0C2}1.06 & \cellcolor[HTML]{C6E0C2}2.6   \\ \hline
\end{tabular}%
}
\label{training-ratio}
\end{table}
\subsection{Analysis of String Length Impact}
\begin{figure}[htbp]
  \centering
  \begin{subfigure}[t]{0.42\linewidth}
    \centering
    \includegraphics[width=\linewidth]{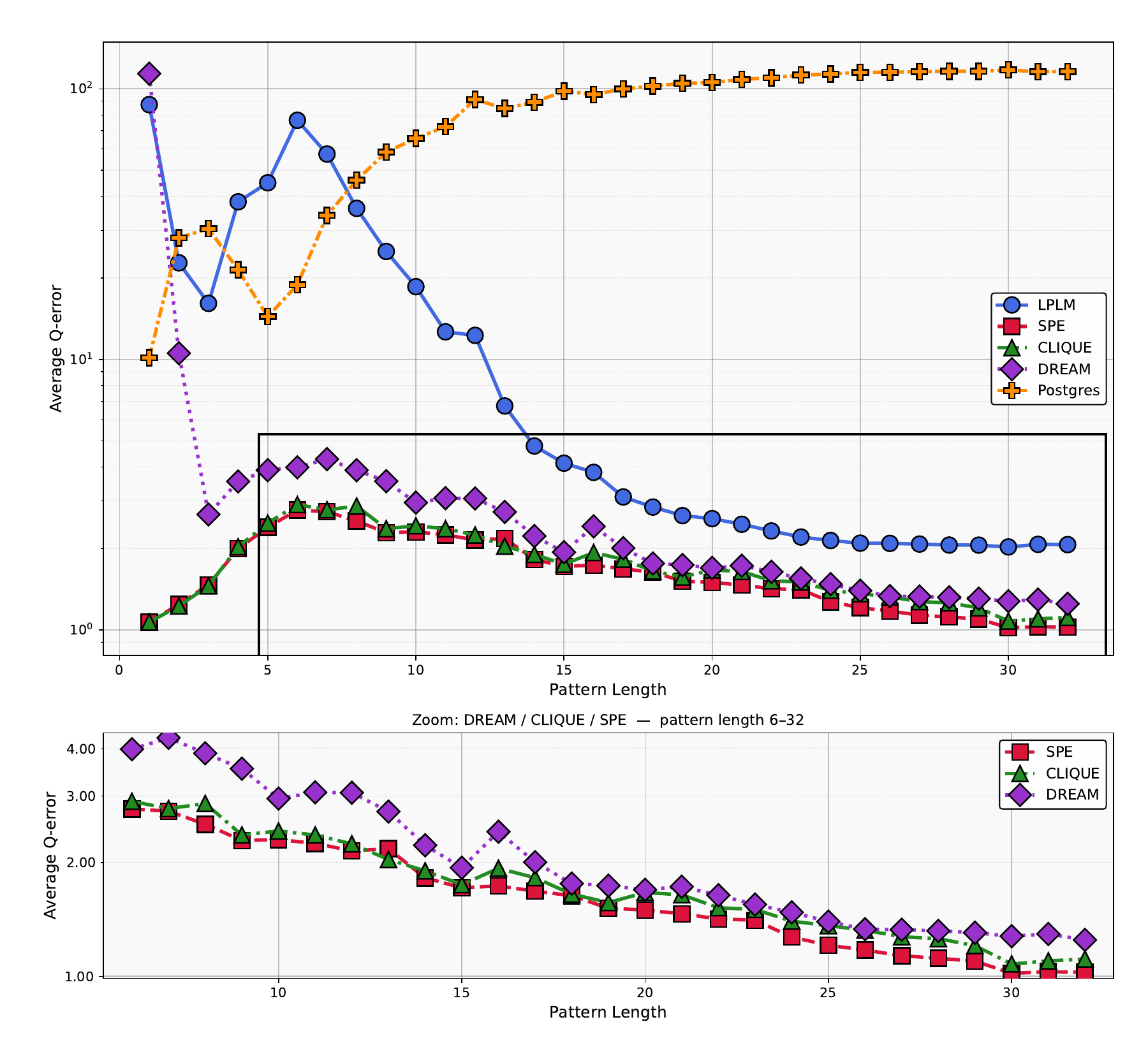}
    \caption{DBLP. SPE maintains low error across all lengths, while CLIQUE, LPLM, and DREAM degrade noticeably for longer patterns.}
    \label{fig:dblp-comparison}
  \end{subfigure}\hfill
  \begin{subfigure}[t]{0.42\linewidth}
    \centering
    \includegraphics[width=\linewidth]{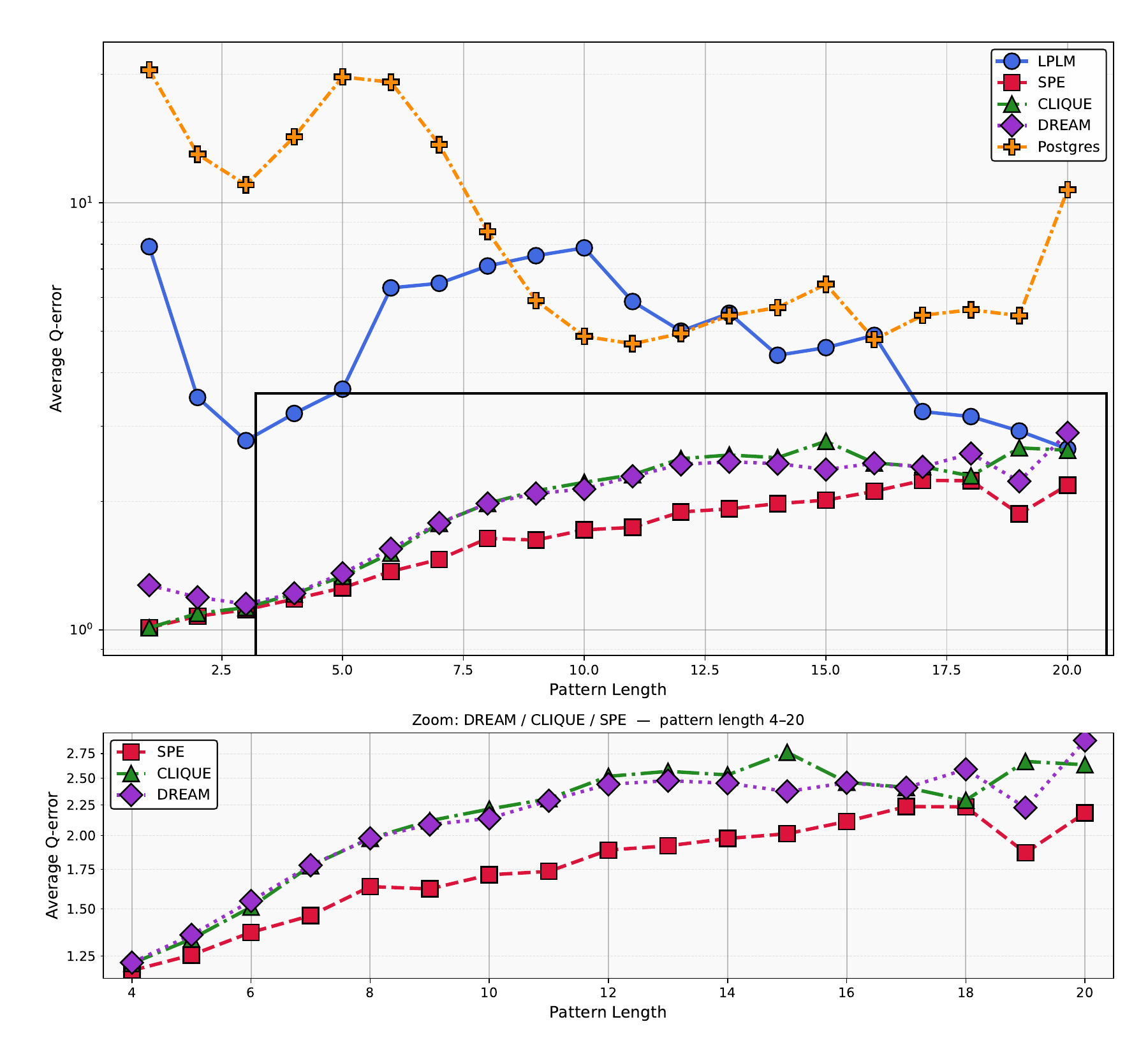}
    \caption{TPCH. SPE achieves the lowest error across all evaluated lengths, with a particularly strong advantage at medium and long pattern lengths.}
    \label{fig:tpch-comparison}
  \end{subfigure}
  \vspace{0.5em}
  \begin{subfigure}[t]{0.42\linewidth}
    \centering
    \includegraphics[width=\linewidth]{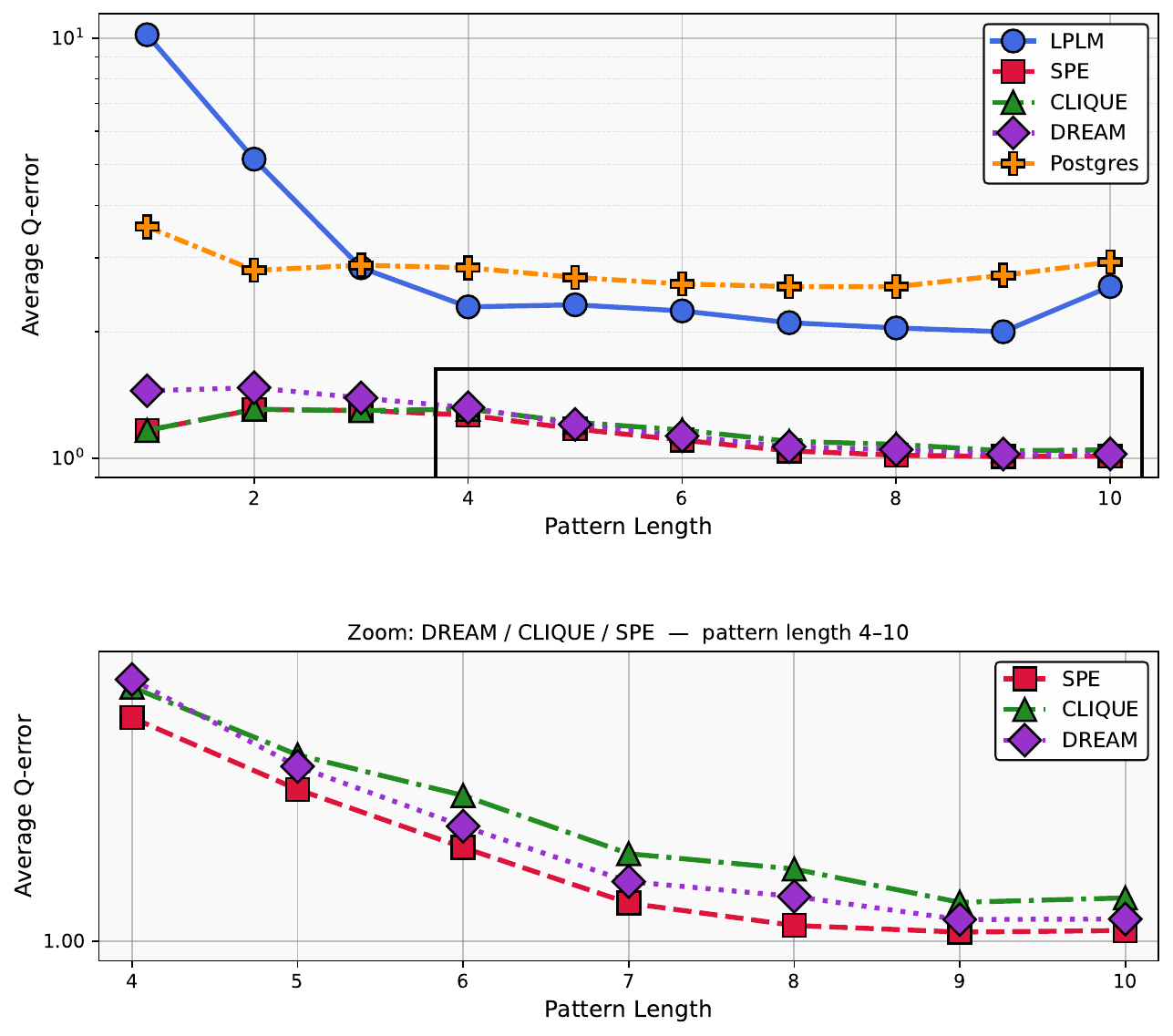}
    \caption{ERGASTF1. All methods perform competitively on this small-vocabulary dataset; SPE achieves the best median Q-error while remaining within a narrow margin of CLIQUE at the tail.}
    \label{fig:ergastf1-comparison}
  \end{subfigure}\hfill
  \begin{subfigure}[t]{0.42\linewidth}
    \centering
    \includegraphics[width=\linewidth]{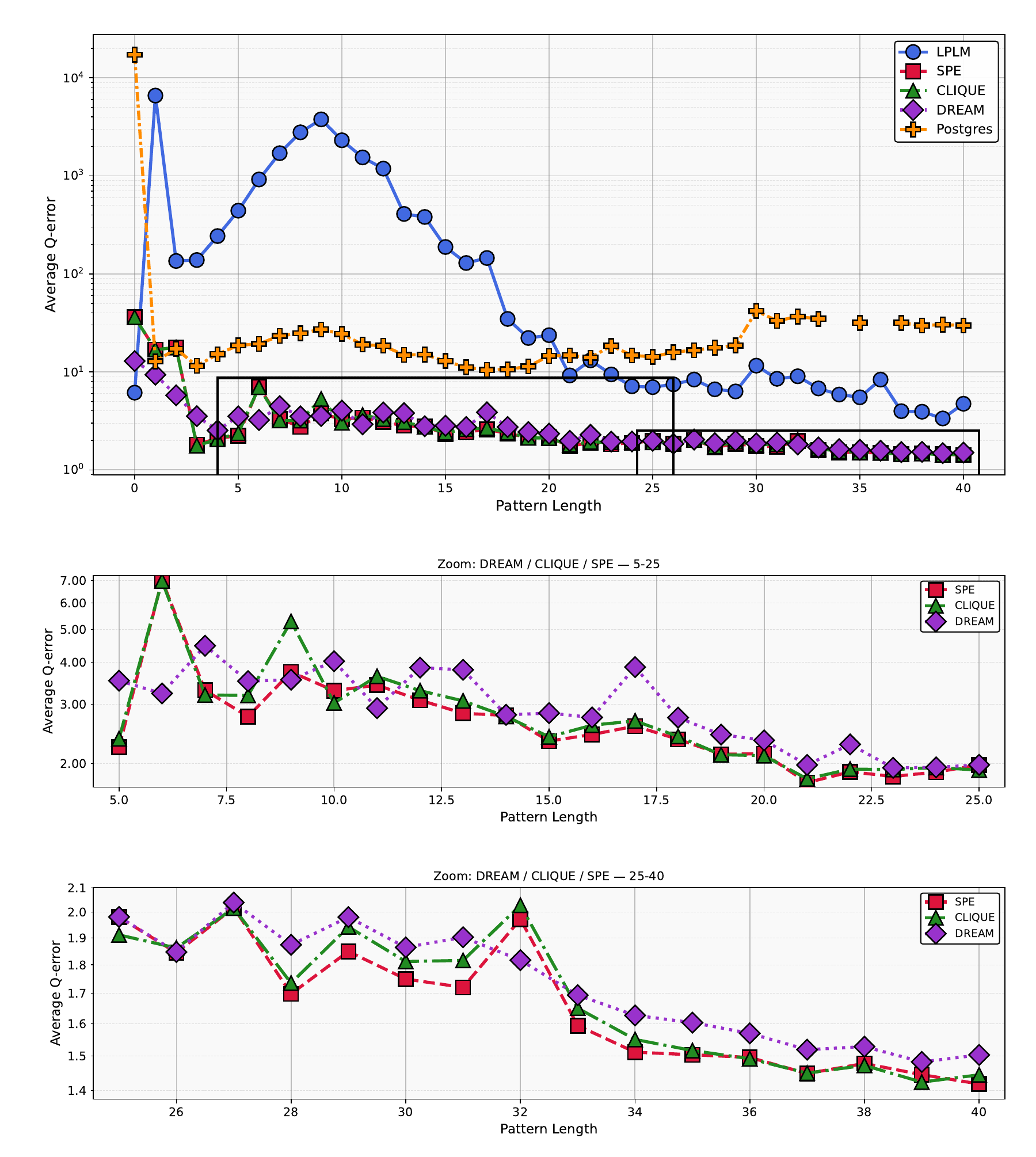}
    \caption{WIKI, which contains long strings (up to 300 characters). SPE's hybrid LSTM-Transformer design sustains low error at greater lengths where purely LSTM-based methods such as CLIQUE and DREAM exhibit higher variance.}
    \label{fig:wiki-comparison}
  \end{subfigure}
  \caption{Average Q-error vs.\ string length on four datasets.}
  \label{fig:length-comparison}
\end{figure}

To compare the behavior of SPE with that of the state-of-the-art model, the average $Q$-error was evaluated across different string lengths (Figure~\ref{fig:length-comparison}). Since real-world applications involve strings of varying lengths, this analysis is particularly important. The results show that SPE achieves the lowest average $Q$-error across most evaluated lengths and datasets.

\subsection{End-to-end execution time}
Q-error measures estimation accuracy in isolation, but it does not reveal
whether a more accurate estimate actually yields a better query plan. Following
the evaluation methodology of ~\cite{leis2015good}, we therefore also
report end-to-end execution time. For each query, the cardinality predicted by a
given estimator is injected into the PostgreSQL optimizer in place of its own
estimate, so that the plan is selected on the basis of that estimator's
prediction; the resulting plan is then executed and the total runtime is
measured. Under this protocol, all methods run on the same engine and the same
data, and the only source of variation is the injected cardinality. TrueCard
denotes the oracle setting in which the exact cardinality is injected, and thus
serves as a lower reference point for what any estimator can achieve. Table~\ref{tab:e2e-single-attr}
reports the results for string-only queries and Table~\ref{tab:e2e-multi-attr}
for multi-attribute queries.
\begin{table}[htbp]
\centering
\caption{End-to-end execution time for single-attribute queries. (The best results are highlighted in dark green, while the second-best results are highlighted in light green)}
\setlength{\tabcolsep}{2pt}
{\small%
\begin{tabular}{|c|c|ccccc|}
\hline
                           &                           & \multicolumn{5}{c|}{End-to-End Execution Time (ms)}                                                                                                                   \\ \cline{3-7}
\multirow{-2}{*}{Datasets} & \multirow{-2}{*}{Methods} & 50th                            & 95th                           & 99th                            & AVG                            & MAX                             \\ \hline
                           & CLIQUE                    & 501.14                          & \cellcolor[HTML]{EFF4ED}808.29 & 1731.59                         & 474.75                         & 3197.17                         \\
                           & DREAM                     & 501.46                          & 810.72                         & 1731.63                         & \cellcolor[HTML]{EFF4ED}474.70 & \cellcolor[HTML]{C6E0C2}2955.83 \\
                           & PostgreSQL                & 502.52                          & 821.26                         & 1798.08                         & 479.74                         & 2973.87                         \\
                           & LPLM                      & \cellcolor[HTML]{EFF4ED}500.23 & 810.89                         & \cellcolor[HTML]{EFF4ED}1662.26 & 474.97                         & 3011.42                         \\
                           & SPE (ours)                & \cellcolor[HTML]{C6E0C2}497.61  & \cellcolor[HTML]{C6E0C2}804.75 & \cellcolor[HTML]{C6E0C2}1583.18 & \cellcolor[HTML]{C6E0C2}469.51 & \cellcolor[HTML]{EFF4ED}2959.90 \\
\multirow{-6}{*}{WIKI-S}     & True Card                  & 501.01                          & 803.96                         & 1596.40                         & 473.55                         & 3049.59                         \\ \hline
                           & CLIQUE                    & 580.25                          & \cellcolor[HTML]{EFF4ED}660.78 & \cellcolor[HTML]{EFF4ED}691.41  & \cellcolor[HTML]{EFF4ED}552.73 & \cellcolor[HTML]{EFF4ED}800.46  \\
                           & DREAM                     & 580.59                          & 661.18                         & 699.98                          & 553.29                         & 831.84                          \\
                           & PostgreSQL                  & 580.68                          & 661.55                         & 698.67                          & 553.68                         & 865.27                          \\
                           & LPLM                      & \cellcolor[HTML]{EFF4ED}580.12  & 661.03                         & 693.04                          & 552.97                         & 824.50                          \\
                           & SPE (ours)                & \cellcolor[HTML]{C6E0C2}579.68  & \cellcolor[HTML]{C6E0C2}657.40 & \cellcolor[HTML]{C6E0C2}681.50  & \cellcolor[HTML]{C6E0C2}551.44 & \cellcolor[HTML]{C6E0C2}797.18  \\
\multirow{-6}{*}{DBLP-S}     & True Card                  & 580.45                          & 660.98                         & 690.91                          & 552.76                         & 796.02                          \\ \hline
                           & CLIQUE                    & \cellcolor[HTML]{EFF4ED}108.59  & \cellcolor[HTML]{EFF4ED}412.74 & 467.12                          & \cellcolor[HTML]{EFF4ED}157.24 & \cellcolor[HTML]{EFF4ED}523.45  \\
                           & DREAM                     & 108.75                          & 413.71                         & 471.45                          & 157.31                         & 544.24                          \\
                           & PostgreSQL                  & 115.44                          & 432.48                         & 485.86                          & 165.85                         & 561.43                          \\
                           & LPLM                      & 108.68                          & 415.20                         & \cellcolor[HTML]{EFF4ED}466.38  & \cellcolor[HTML]{EFF4ED}157.24 & 533.33                          \\
                           & SPE (ours)                & \cellcolor[HTML]{C6E0C2}98.75   & \cellcolor[HTML]{C6E0C2}388.33 & \cellcolor[HTML]{C6E0C2}432.86  & \cellcolor[HTML]{C6E0C2}148.28 & \cellcolor[HTML]{C6E0C2}503.65  \\
\multirow{-6}{*}{TPCH-S}     & True Card                  & 106.02                          & 416.94                         & 462.72                          & 157.26                         & 531.11                          \\ \hline
                           & CLIQUE                    & 0.58                            & \cellcolor[HTML]{EFF4ED}0.66   & \cellcolor[HTML]{EFF4ED}0.70    & \cellcolor[HTML]{EFF4ED}0.56   & 0.84                            \\
                           & DREAM                     & 0.60                            & 0.67                           & \cellcolor[HTML]{EFF4ED}0.70    & \cellcolor[HTML]{EFF4ED}0.56   & 0.75                            \\
                           & PostgreSQL                  & 0.61                            & 0.68                           & 0.80                            & 0.58                           & 0.76                            \\
                           & LPLM                      & \cellcolor[HTML]{EFF4ED}0.57    & 0.67                           & \cellcolor[HTML]{EFF4ED}0.70    & \cellcolor[HTML]{EFF4ED}0.56   & \cellcolor[HTML]{EFF4ED}0.73    \\
                           & SPE (ours)                & \cellcolor[HTML]{C6E0C2}0.56    & \cellcolor[HTML]{C6E0C2}0.64   & \cellcolor[HTML]{C6E0C2}0.66    & \cellcolor[HTML]{C6E0C2}0.54   & \cellcolor[HTML]{C6E0C2}0.72    \\
\multirow{-6}{*}{ERGASTF1-S} & True Card                  & 0.59                            & 0.66                           & 0.70                            & 0.57                           & 0.73                            \\ \hline
\end{tabular}%
}
\label{tab:e2e-single-attr}
\end{table}

\begin{table}[htbp]
\centering
\caption{End-to-end execution time for multi-attribute queries}
\setlength{\tabcolsep}{3pt}
{\small%
\begin{tabular}{|c|c|ccccc|}
\hline
                           &                         & \multicolumn{5}{c|}{End-to-End Execution Time(ms)}                                                                                                                 \\ \cline{3-7}
\multirow{-2}{*}{Datasets} & \multirow{-2}{*}{Model} & 50th                           & 95th                           & 99th                           & AVG                            & MAX                            \\ \hline
                           & MultiCard               & \cellcolor[HTML]{C6E0C2}92.97  & \cellcolor[HTML]{C6E0C2}192.25 & \cellcolor[HTML]{C6E0C2}215.78 & \cellcolor[HTML]{C6E0C2}98.26  & \cellcolor[HTML]{C6E0C2}249.71 \\
                           & PostgreSQL                & \cellcolor[HTML]{EFF4ED}93.25  & \cellcolor[HTML]{EFF4ED}192.61 & \cellcolor[HTML]{EFF4ED}216.09 & \cellcolor[HTML]{EFF4ED}98.55  & \cellcolor[HTML]{EFF4ED}250.02 \\
\multirow{-3}{*}{IMDB-M}     & True Card               & 93.04                          & 192.49                         & 215.59                         & 98.40                          & 249.78                         \\ \hline
                           & MultiCard               & \cellcolor[HTML]{EFF4ED}382.36 & \cellcolor[HTML]{C6E0C2}681.74 & \cellcolor[HTML]{C6E0C2}740.54 & \cellcolor[HTML]{C6E0C2}363.61 & \cellcolor[HTML]{C6E0C2}876.65 \\
                           & PostgreSQL                & \cellcolor[HTML]{C6E0C2}364.60 & \cellcolor[HTML]{EFF4ED}683.57 & \cellcolor[HTML]{EFF4ED}743.35 & \cellcolor[HTML]{EFF4ED}364.60 & \cellcolor[HTML]{EFF4ED}876.87 \\
\multirow{-3}{*}{DBLP-M}     & True Card               & 388.84                         & 681.88                         & 740.39                         & 365.60                         & 877.55                         \\ \hline
                           & MultiCard               & \cellcolor[HTML]{C6E0C2}51.39  & \cellcolor[HTML]{C6E0C2}78.74  & \cellcolor[HTML]{C6E0C2}89.90  & \cellcolor[HTML]{C6E0C2}49.69  & \cellcolor[HTML]{C6E0C2}137.37 \\
                           & PostgreSQL                & \cellcolor[HTML]{EFF4ED}53.27  & \cellcolor[HTML]{EFF4ED}87.10  & \cellcolor[HTML]{EFF4ED}114.04 & \cellcolor[HTML]{EFF4ED}53.00  & \cellcolor[HTML]{EFF4ED}201.35 \\
\multirow{-3}{*}{TPCH-M}     & True Card               & 52.26                          & 81.11                          & 97.14                          & 51.08                          & 166.77                         \\ \hline
                           & MultiCard               & \cellcolor[HTML]{C6E0C2}0.34   & \cellcolor[HTML]{C6E0C2}0.59   & \cellcolor[HTML]{C6E0C2}0.64   & \cellcolor[HTML]{C6E0C2}0.36   & \cellcolor[HTML]{C6E0C2}0.86   \\
                           & PostgreSQL                & \cellcolor[HTML]{EFF4ED}0.42   & \cellcolor[HTML]{EFF4ED}0.65   & \cellcolor[HTML]{EFF4ED}0.73   & \cellcolor[HTML]{EFF4ED}0.40   & \cellcolor[HTML]{EFF4ED}3.02   \\
\multirow{-3}{*}{ERGASTF1-M} & True Card               & 0.32                           & 0.61                           & 0.68                           & 0.37                           & 2.76                           \\ \hline
\end{tabular}%
}
\label{tab:e2e-multi-attr}
\end{table}

\section{Conclusion and Future Work}
This paper makes two key contributions. First, we introduced the SPE model, which can function independently to estimate selectivities for string predicates, consistently outperforming existing approaches across varying string lengths. Second, by integrating the SPE model with numerical filters, we presented MultiCard, a robust end-to-end model designed to accurately estimate real-world workloads involving complex combinations of string and numeric predicates. Future research directions include extending the model's capabilities to support join queries, as well as investigating methods to reduce the model's training overhead while strictly preserving high estimation accuracy.


\FloatBarrier
\bibliographystyle{plainnat}
\bibliography{MultiCard-Reference}

\end{document}